\documentclass[aps,pra,groupedaddress]{revtex4-1}
\usepackage[latin1]{inputenc}
\usepackage{graphicx} 
\usepackage{amsmath}
\usepackage{amsfonts}
\usepackage{amssymb}

\newcommand\bra[1]{\langle\,#1\,\vert}
\newcommand\ket[1]{\vert\,#1\,\rangle}

\def \H{{\cal{H}}}

\def \barl {\begin{array}{rl}} 
\def \ea {\end{array}}

\newcommand \Tr[1]{{\rm Tr}\{ {#1}\}}

\let\abs=\envert

\begin{document}
\title{Mechanisms of irreversible decoherence in solids.}
\author{F.D. Domínguez, R.C. Zamar, H.H. Segnorile, C.E. González}
\affiliation{FaMAF, Universidad Nacional de Córdoba; IFEG, Conicet}
\begin{abstract}

Refocalization sequences in Nuclear Magnetic Resonance (NMR) can in principle reverse the coherent evolution under the secular dipolar Hamiltonian of a closed system. 
We use this experimental strategy to study the effect of irreversible decoherence on the signal amplitude attenuation in a single crystal hydrated salt where the nuclear spin system consists in the set of hydration water proton spins having a strong coupling within each pair and a much weaker coupling with other pairs. 
We study the experimental response of attenuation times with temperature, crystal orientation with respect to the external magnetic field and rf pulse amplitudes. 
We find that the observed attenuation of the refocalized signals can be explained by two independent mechanisms: (a) evolution under the non-secular terms of the reversion Hamiltonian, and (b) an intrinsic mechanism having the attributes of irreversible decoherence induced by the coupling with a quantum environment.
To characterize (a) we compare the experimental data with the numerical calculation of the refocalized NMR signal of an artificial, closed spin system. To describe (b) we use a model for the irreversible adiabatic decoherence of spin-pairs coupled with a phonon bath which allows evaluating an upper bound for the decoherence times. This model accounts for both the observed dependence of the decoherence times on the eigenvalues of the spin-environment Hamiltonian, and the independence on the sample temperature. This result, then, supports the adiabatic decoherence induced by the dipole-phonon coupling as the explanation for the observed irreversible decay of reverted NMR signals in solids.

\end{abstract}
\maketitle

\section{Introduction}
A cutting-edge subject in the research of irreversible processes is the study of the quantum dynamics of many-particle interacting systems  coupled with a quantum correlated environment. In this scenario quantum decoherence represents a fascinating problem with links to fundamentals, as well as to modern application fields. 
In the first area, decoherence is considered by many as the mechanism responsible for the emergence of the classical world from the microscopic quantum mechanical world \cite{joos2003decoherence,zurek03}. In the other,  implementation of applications like scalable quantum registers, demands handling collective coherent states. The collective coherence that can be prepared in a multiparticle cluster becomes fragile due to coupling with the environment, turning it crucial to getting insight on the subtle mechanism by which the coherence loss occurs. Such process, which involves no energy exchange with the outworld, is called adiabatic quantum decoherence \cite{yu-eberly02}.

Nuclear magnetic resonance (NMR) of spin ensambles in the solid state can serve as a suitable test bed for the quantum dynamics on large clusters of interacting particles. In fact, NMR provides a variety of techniques to create and manipulate coherent spin states. Particularly, a class of refocusing (often called `reversion') experiments allows retrieving the multi-spin dynamics governed by the dipole-dipole interaction of an ideal {\em closed}  spin system. Actual experiments, however, yield refocused signals whose amplitudes attenuate with the reversion time.  The source of such attenuation can be connected to experimental causes, and, from a microscopic viewpoint, to the unavoidable coupling of the observed system with other degrees of freedom  \cite{PetitjeanJacq06,GzzSegZam11,SegZam11}. 
In this work we use reversion experiments to isolate these effects by monitoring the signal attenuation times as a function of controlled experimental variables: efficiency of the reversion pulse sequences, sample temperature, and orientation with respect to the external magnetic field.

The explanation of irreversible decoherence in solid state NMR, and the role played by the environment in this process has remained as an open question for a long time \cite{morgan2012multispin}. Particularly, the mechanism by which nuclear spins are able to achieve a state of quasi-equilibrium continues to be elusive nowadays. The NMR literature seldom relates the signal attenuation in reversion experiments in solids with the environment-induced destruction of the coherent superposition of states (environmental decoherence). 
A possible reason for this is the fact that strongly interacting spin systems, like solids, are well isolated of the environment degrees of freedom (thermal fluctuations), which manifests as `very long' spin-lattice relaxation times. This generally leads to expecting negligible effect of the environment on the spin dynamics over the earlier timescale. 
However, the fact that spin-lattice mechanisms in solids cannot account for the irreversible decay of the NMR signal does not rule out the occurrence of  quantum decoherence, which involves the loss of local phases, within a time regime where the spin-lattice energy exchange is still ineffective. 
On the contrary, in the field of open quantum systems, it is currently accepted that the many-body character of the observed systems is a decisive condition for the occurrence of system-environment correlation, associated with the entanglement and quantum decoherence of the observed system \cite{priv98,SegZam11,yu-eberly02,strunz16,haikka2013}.  Such mechanism has a characteristic timescale much shorter than that of thermalization or spin-lattice relaxation, and can be thought of as a main microscopic source of the signal decay in reversion experiments and also in the occurrence of quasi-equilibrium (or pseudo-thermalization) states \cite{SegZam11,GzzSegZam11,dominguez2016irreversible}. 

A recent theoretical proposal describes the irreversible adiabatic decoherence of a system of weakly interacting spin pairs coupled with a phonon field \cite{dominguez2016irreversible}. 
The model considers that all spins in the sample are part of a complex dipolar network while the system-environment interaction is the fluctuation in the strongest dipolar couplings due to the low frequency phonons. 
The resulting time dependence of the reduced density matrix elements is a product of the corresponding element of an isolated spin system and a decoherence function which introduces an irreversible decay. The rate of this decay increases with the Hamming distance between the involved states and \emph {with the intrapair dipolar coupling intensity}.
This coherence loss can be reflected in the decay of the expectation values that represent the system observables, and consequently on the measured signal amplitude.

In this work we use the well known `magic echo' NMR reversion sequence (ME) on a single crystal sample of CaSO4·2H2O (di-hydrated calcium sulphate or gypsum), as a good representation of the ideal system treated in \cite{dominguez2016irreversible}. Our goal is to expose the main causes that attenuate the echo signal. 
We analyze two mechanisms very different in nature: the experiment non-ideality, predominantly given by the evolution under non-secular terms of the reversion Hamiltonian, and the environment induced adiabatic decoherence. 

The decay time dependence on the intrapair dipolar frequency and temperature is contrasted with theoretical estimates. To quantify the role of the non-secular terms we compute their effect on the signal amplitude attenuation on an artificial ten-spin closed system having the same symmetry and orientations than the measured sample.  
The contribution of decoherence is interpreted in terms of predictions from the theory of open quantum systems, using the variation of the system {\em purity}  as a quantifier of decoherence \cite{ZurekHabibPaz93}. On this basis we calculate the purity, using the reduced density matrix from the pairs-phonon model  \cite{dominguez2016irreversible} and a reasonable hypothesis for the multispin correlation growth in the dipolar network.  The purity also is an upper bound for the observable NMR signal, then, comparison of the calculated purity rates with the experimental attenuation times allows to conclude that the pair-phonon model provides a proper explanation to the signal loss in the solid state NMR reversion experiments and consequently to the development of quasi-equilibrium states.

\section{The sample}

\begin{figure}
\includegraphics[width=5.73cm,height=5.48cm]{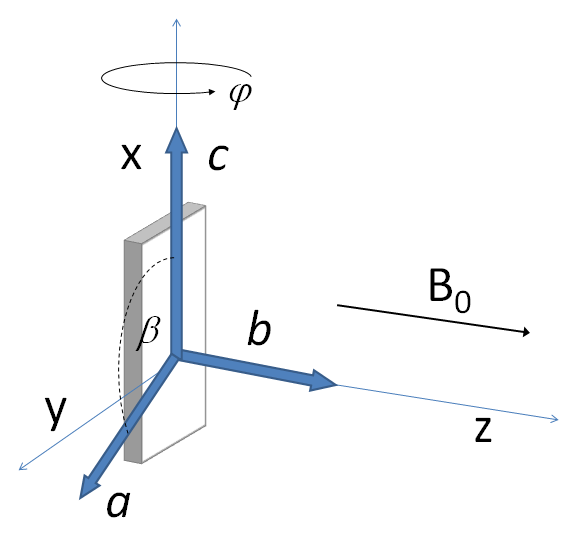}
\caption{Orientation of the single crystal with respect to the external magnetic field used in the experiment. The sample can be rotated around axis $c$. }
\label{dir_cristal}
\end{figure}

Gypsum (CaSO$_4\cdot$2H$_2$O) is a paradigmatic example of a hydrated salt, whose crystal structure has been resolved long ago and was redetermined more recently \cite{boeyens2002gypsum}. The unit cell is monoclinic, with $a,b,c = 6.28$\AA, $15.20$\AA, $6.52$\AA, and $\beta = 127.4°$. The sample used in this study is a piece of natural, transparent gypsum single crystal, with the size ($A$=10 $\times$ $B$=2  $\times$ $C$=12 mm$^3$), where $A, B$ and $C$ are parallel to the primitive cell axes $a, b$ and $c$, respectively. It was placed on a holder which allows rotating the crystal around the $c$ axis, perpendicular to the external field $\vec{B}_0$, as shown in Fig.\ref{dir_cristal}. Let us call $\varphi$ the angle between axis $b$ and $\vec{B}_0$, and define $\varphi =0$ when both vectors are parallel. 

Gypsum has only one observable spin species: the hydration water protons, which adopt two different orientations. The geometry of this lattice of regularly distributed spin pairs entails a hierarchy of dipolar interactions which can roughly be grouped into stronger (mainly produced by the intrapair interactions) and weaker. The NMR spectrum of this arrangement shows a splitting $\omega_D $ (Pake doublet)  which depends, to first order, on the magnitude of the dipolar intrapair interaction and consequently on the orientation of the sample with respect of the Zeeman field \cite{pake48}.  In our experiment we take advantage of this fact to manipulate $\omega_D$ by rotating the sample around a direction  perpendicular to $\vec B_0$.

\begin{figure}
\includegraphics[width=8.15cm,height=10.30cm]{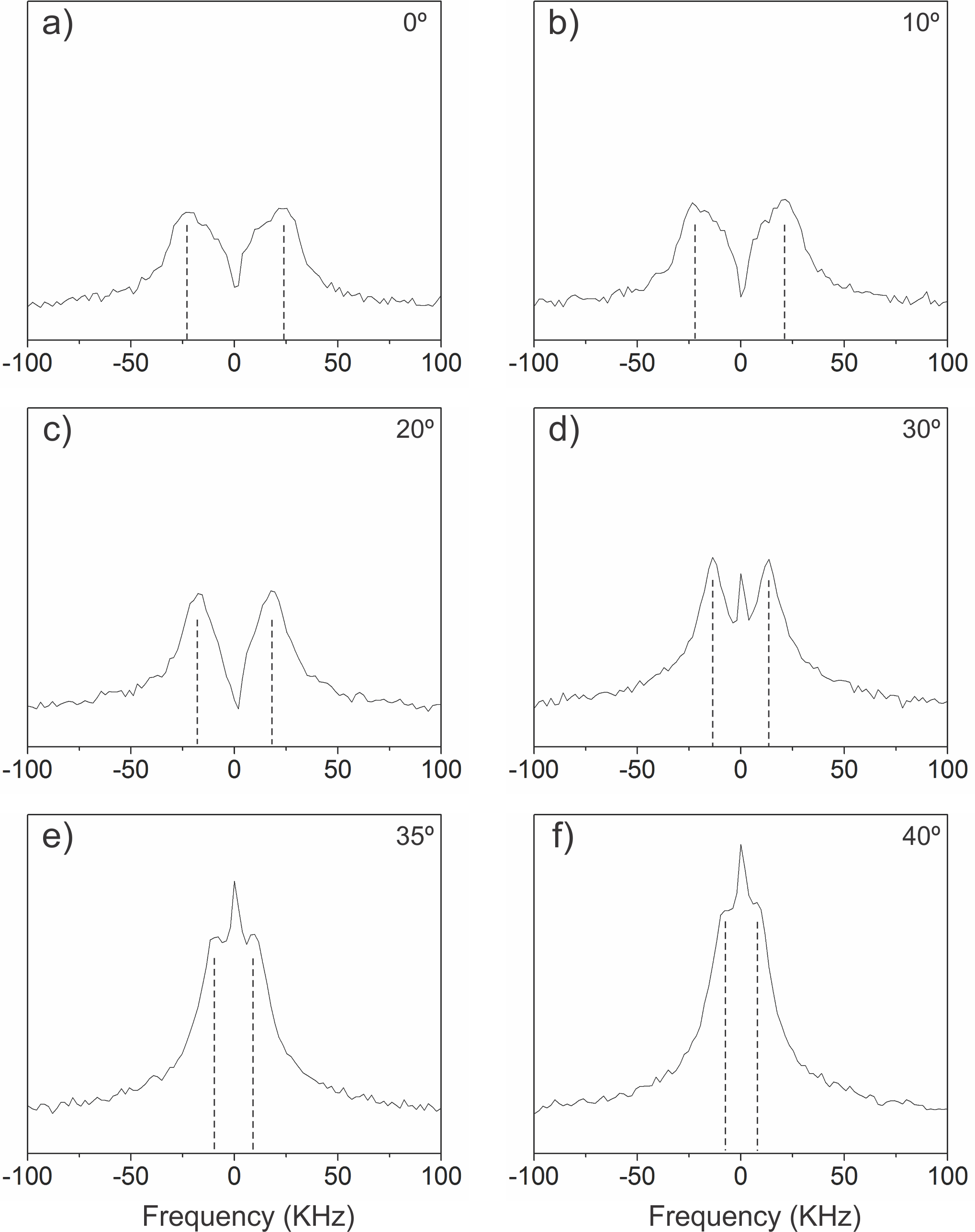}
\caption{Spectra of a gypsum single crystal with axis $c$ perpendicular to the magnetic field and axis at six orientations of axis $c$ respect to the magnetic field. The spectra show resolved doublets for $\varphi= 0, 10, 20$ degrees, while those corresponding to  $\varphi= 30, 35, 40$ degrees, also have a central peak.
}
\label{spectra}
\end{figure}

 	\begin{table}\label{angles}	
 	\begin{center} 
		\begin{tabular}{ | c | c |c |}  
			\hline  
			$\varphi (º)$ & $\omega_D$ $\pm 2$ ($\mathrm{KHz}/2\pi$)& $M_2$ $\pm 5$ ($\mathrm{KHz}^2$) \\
			\hline 
			0 & 46 & 1320 \\ \hline
			10 & 42 &  1170\\ \hline
			20 & 36 & 990 \\ \hline
			30 & 26 & 620\\ \hline
			35 & 18  & 315\\ \hline
			40 & 15 & 170\\
			\hline 
		
		\end{tabular} 
	\caption{Experimental dipolar splitting $\omega_D$ for each crystal orientation $\varphi$ and the respective second moment $M_2$ of the NMR spectra of gypsum.} 
	\end{center}
	
	\end{table}
 The spectra corresponding to the six different orientations and all the experiments shown in this work  were measured  on a Bruker minispec mq-20 spectrometer, at 20 MHz. The six spectra shown in Fig. \ref{spectra}, were fitted with gaussian functions; the relation $\omega_D(\varphi)$ was determined from the dipolar splitting of the spectra and is shown in the second column of table I. 
Spectra corresponding to angles $\varphi= 0$ and $10$ have two clearly resolved, asymmetric peaks, as expected in a system of weakly interacting spin pairs \cite{keller88}.  The tabulated frequencies correspond to the vertical lines at $\omega_D = \pm 23$ and $\pm 21$ kHz shown in Fig. \ref{spectra} a) and b) respectively. The spectrum of $\varphi= 20º$ has resolved symmetric peaks. The spectra from orientations $\varphi= 30$, 35 and 40 degrees are more complex because the water molecules are clearly not equivalent; the central peak seen on Fig. \ref{spectra} d), e) and f) corresponds to one kind of molecules and the resolved doublets to the other. Again, we considered the splitting of the largest frequency spectra  to define  $\omega_D$. 
The rigth column of table I shows the second moments,  $M_2$, of the gypsum spectra obtained as the sum of the intra-pair ($\omega_D^2/4\pi^2$) and inter-pair (width of each doublet component)  contributions. 

\section{Measurement of decoherence}	

\subsection{The magic echo sequence}
\begin{figure}
\includegraphics[width=8cm,height=2.08cm]{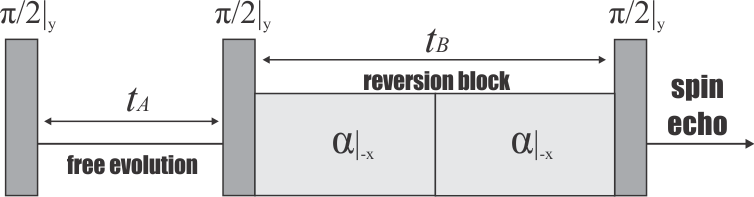}
\caption{Pulse sequence used in the experiment. The block of duration $t_B$ reverts the evolution under the secular dipolar Hamiltonian which takes place during $t_A$.}
\label{MEsequence}
\end{figure}

We use the well known radio frequency pulse sequence called {\em magic echo} (ME) \cite{rpw71}, which has the effect of refocusing the time evolution due to the secular dipolar interaction of dipolar coupled spin systems in a large magnetic field. The ME sequence is shown in Fig.\ref{MEsequence}. The reversion block is composed of two pulses of amplitude $\omega_{1}$ (in frequency units), length $\alpha$, and alternating phases $x,-x$, which are ``sandwiched'' by two $\frac{\pi}{2}$ hard pulses of phases $y$ and $-y$. Alternating phases $x,-x$ have the effect of avoiding accumulation of phase errors \cite{rpw71}. 

The effect of this block can be clearly shown by considering a simplified sequence with only one pulse with phase $x$, duration $t_B$ and intensity $\omega_1$ inside the sandwich. The corresponding propagator is 
\begin{equation} 
U_{ME}= R_y(\frac{\pi}{2})\exp{\left[ -i \; t_{B} \left(\omega_{1}I_x+\; \H_D^0 \right)\right]\; R_y(-\frac{\pi}{2})},
\end{equation}
where $R_y(\beta)= e^{-i\beta I_y}$ represent the hard pulses and $\mathcal{H}_D^0 = \sum_{k,j} \sqrt{1/6}\omega_D^{kj} T_{20}^{kj} $ is the total secular dipolar energy.  $ T_{20}^{kj}=2/\sqrt{6}\left( 3I_z^{(k)}I_z^{(j)}-\mathbf{I}^{(k)} \cdot\mathbf{I}^{(j)}\right) $ is the irreducible tensor operator corresponding to spins $k$ and $j$.
Using the relations  
\begin{equation} \begin{array}{l}
 Ae^BA^{-1}=e^{ABA^{-1}} ,\\
 e^{-i\frac{\pi}{2}I_y}I_xe^{i\frac{\pi}{2}I_y}=-I_z,\\
 e^{-i\frac{\pi}{2}I_y}T_{20}^{k,j}e^{i\frac{\pi}{2}I_y}=-\frac{1}{2}T_{20}^{k,j}+\sqrt{\frac{3}{8}}\left( T_{22}^{k,j}+T_{2-2}^{k,j} \right) , 
\end{array} 
\end{equation}
where $T_{2 \pm2}^{k,j}=2I_{\pm}^{(k)}I_{\pm}^{(j)}$,
we have
\begin{equation} \label{USMcompleto}
 U_{ME}=e^{-i\omega_{1} t_{B}\left\lbrace - I_z+\sum_{k,j}\sqrt{\frac{1}{6}}\frac{\omega_D^{k,j}}{\omega_{1}}\left[ -\frac{1}{2}T_{20}^{k,j}+\sqrt{\frac{3}{8}}\left( T_{22}^{k,j}+T_{2-2}^{k,j} \right)  \right]\right\rbrace   }.
\end{equation}
If the rf intensity $\omega_1$ is bigger than the intrapair dipolar interaction $\omega_{D}$ (the main interaction in a spin pair system)
\begin{equation} \label{rf_gg_D}
\omega_{1}\gg \omega_D \geq\omega_{D}^{k,j},
\end{equation}
one can within first order perturbation theory, disregard the evolution caused by the non-secular term $(T_{22}+T_{2-2})$ \cite{abragol82,rpw71}. Since $[T_{20}, I_z]=0$ it is usual to set the experiment such that 
\begin{equation}
U_{ME}^0\simeq \exp{\left[ +i \; t_{B} \frac{1}{2} \mathcal{H}_D^0 \right]}.
\label{UMEsecular}
\end{equation}
This expression brings about the main benefit of the ME, which is to revert the sign of the time evolution under the secular dipolar Hamiltonian and then ``undo'' one-half of its action.
The typical experiment starts with the spin system in thermal equilibrium $\rho(0^-)\propto I_z$. The first, saturating $\left(\frac{\pi}{2}\right)$, rf. pulse of phase $y$ at time $t=0$ leaves  $\rho(0^+)\propto I_x$. The system evolves freely (under $\H_D^0$) during $t_A$ and backwards during $t_B$. Selecting $t_B= 2t_A$ will ideally revert the evolution to its state at $t=0$, and the NMR signal $\Tr{\rho I_x}$ should recover its initial amplitude. 

The simple pulse sequence described above contains the essence of the reversion methods. Of course, there are various uncontrollable experimental settings that may overshadow the ideal response.  There are many other ingenious sequences based on ME which combine more ME modules with different phases that prevent or at least mitigate some of the various possible non-idealities of the actual experiment.
It is clear, however, that the efficiency of the reversion sequence is restricted by condition  (\ref{rf_gg_D}), and that the experimental realization of this condition may not be possible for strong dipolar couplings. This may cause that part of the observed time evolution be due to the non-secular term, which may enhance the degradation of the observed echoes.

The purpose of this work is to isolate the different mechanisms that attenuate the reverted echo in a reversion experiment. Particularly, we look for a sign of the adiabatic decoherence due to the coupling of the spin system with the environment.

\subsection{Experimental Results} \label{expresult}

The signal amplitude recorded at the end of the ME sequence (see Fig.\ref{MEsequence}) attenuates as a function of the elapsed time $t_A$ (and consequently of time $t_B$ also) with a characteristic time $T_M$ at which the signal amplitude reduces to $1/e$ of its initial value. 
According to Eq.(\ref{USMcompleto}), one can expect the efficiency of the reversion sequence to depend on both, the intrapair frequency  $\omega_{D}$ (Table I) and the radiation intensity $\omega_1$, therefore we measured the attenuation time $T_M$ of the reverted signals for different values of  $\omega_{1}$ and also varied  $\omega_D$ by changing the sample orientation respect to the external field. 

\begin{figure}
\includegraphics[width=7.65cm,height=6.24cm]{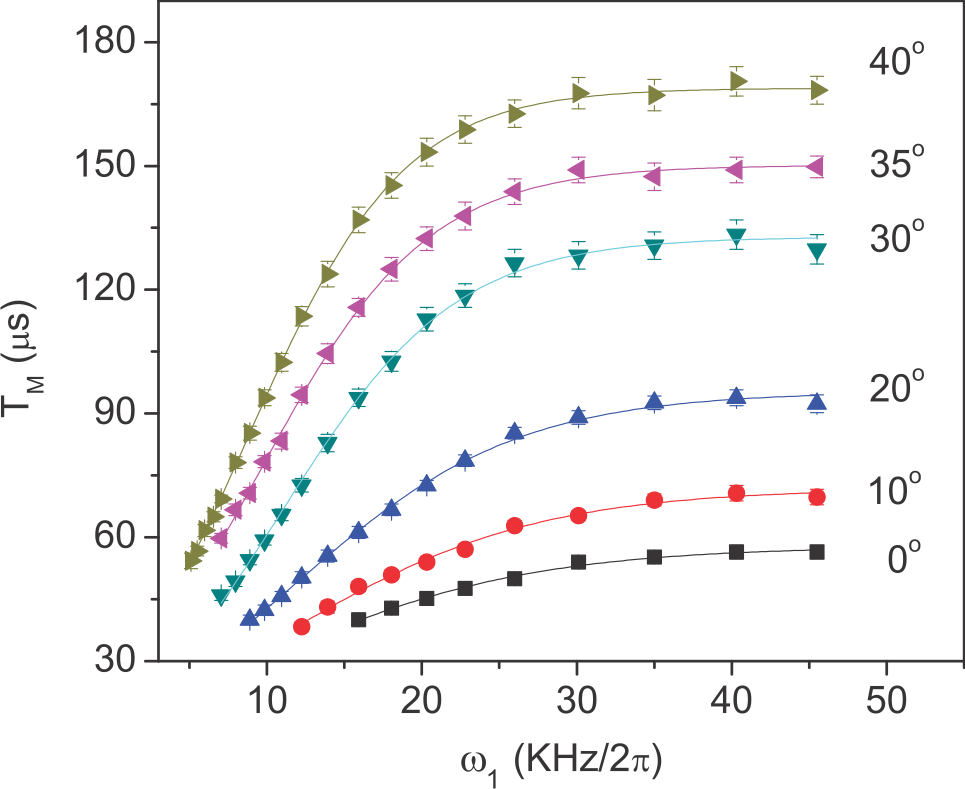}
\caption{Attenuation times $T_M$ of the reverted signal amplitudes as a function of the irradiation intensity $\omega_{1}$ of the ME sequence. Different curves (symbols) correspond to each studied orientation of the single crystal sample with respect to the external magnetic field. Solid lines are fittings to a sigmoid $A(1+e^{-C(\omega_1+B)})^{-1}$.}
\label{tauvsnu1}
\end{figure}

 Figure \ref{tauvsnu1} shows the dependence of $T_M$ with $\omega_{1}$. Each curve corresponds to a fixed crystal orientation, then, to a given value of $\omega_D$. The experiment was performed at  $T=220$ K and 310 K, and the obtained attenuation curves were noticeably  independent on the sample temperature, since the corresponding $T_M$ are identical within the experimental error. 

The fact that the attenuation times are very short for small $\omega_{1}$ at all crystal orientations, is consistent with a poor reversion efficiency for low values of $\omega_{1}/\omega_D$, as follows from Eq.(\ref{USMcompleto}). All the data curves rise with an approximately linear trend whose slope depends on the sample orientation. The salient characteristic is that all the curves show a plateau at higher values of $\omega_{1}$, which implies that the reversion efficiency cannot be improved by increasing the amplitude of the alternating phase pulses in the ME sequence. The maximum values of $T_M$ attained by the different curves have a marked dependence on  $\omega_D$.
In fact, the data are adequately fitted (solid lines in Fig.\ref{tauvsnu1}) by sigmoids of the form
\begin{equation} \label{sigmoide}
T_M=\frac{A}{1+e^{-C(\omega_{1}+B)}}=\left( \frac{e^{-CB}}{A}\frac{1}{e^{C\omega_1}}+\frac{1}{A}\right)^{-1} ,
\end{equation}
which strongly suggests that the signal loss is dominated by two different process: one that depends exponentially on $\omega_{1}$, and another that is independent of $\omega_{1}$. We associate the former with the evolution under the non-secular terms of Eq.(\ref{USMcompleto}) that are not reverted with the MS sequence, and the latter with a different process, independent of the experimental settings. 

\subsection{Effect of the non-secular terms} \label{nonsec}

Data from Fig. \ref{tauvsnu1} measured in a wide range of dipolar couplings and radiation intensities, allowed isolating the different sources of decay. Following the reasoning after Eq.(\ref{sigmoide}), we assume that the two attenuation sources are independent and write the measured attenuation times $T_M$  
\begin{equation}
\label{sum_inv}
\frac{1}{T_M} =  \frac{1}{T_{NS}}+ \frac{1}{\tau_D},
\end{equation}
where $1/T_{NS}$ is the decay rate due to the non-secular part of the ME operator $U_{ME}$ in Eq.(\ref{USMcompleto}) and  $1/\tau_D$ is the decoherence rate. 
Other sources of signal decay, as inhomogeneity of rf field or finite width of the hard, rf pulses, are optimized so they become negligible (besides, they are independent of $\omega_1$ and $\omega_D$).

\begin{figure}
	\includegraphics[width=7.59cm,height=6.14cm]{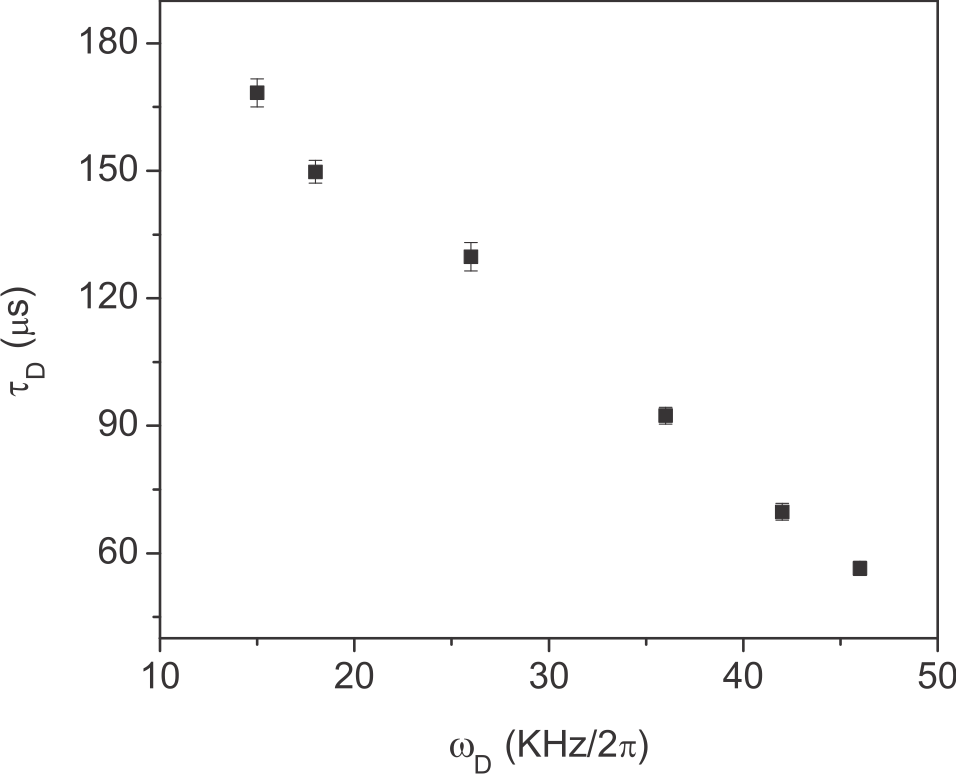}
	\caption{ Maximum attenuation times of reverted signals vs. the dipole frequency splitting of the  NMR spectra corresponding to the six crystal orientations $\varphi=0,10,20,30,35,40$ degrees.}
	\label{tau-wD}
\end{figure}

In order to try this hypothesis we follow two steps. First, it is worth to notice that $\tau_D$, which represents the coupling between the spin system and the environment, does not depend on an experimental parameter as  $\omega_1$, and also that the rate $1/T_{NS} \rightarrow 0$ if $\omega_1 \rightarrow \infty$. This allows, in principle, to identify $\tau_D$ with the plateaus (maximum attenuation time)  of the measured curves. Fig. \ref{tau-wD} shows that $\tau_D$ decreases for increasing vaules of  $\omega_D$ in an approximately linear trend. We notice that $\tau_D$ may be slightly  underrated for the angles $\varphi=0º$ and 10º, because the quotient $\omega_{1}/\omega_D$ is low within the available rf intensity range. 

\begin{figure}
\includegraphics[width=7.65cm,height=6.3cm]{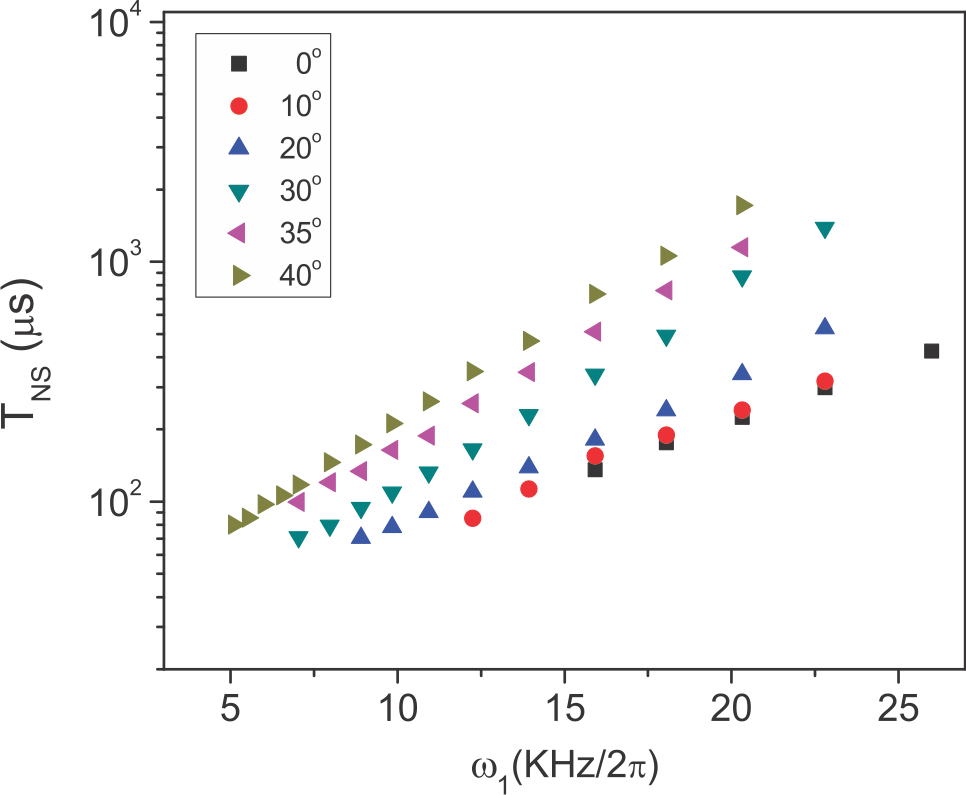}
\caption{Decay times $T_{NS}$ obtained from experimental data using Eq. \ref{sum_inv} within the region where the  dependence on $\omega_1$ is dominated by the non-reverted terms of the ME propagator.}
\label{tns_exp}
\end{figure}

 Having determined $\tau_D$, we calculate the dependence of the decay time $T_{NS}$ on $\omega_1$.  The results are shown in Fig. (\ref{tns_exp}). The linearity of the plotted data indicates that $T_{NS}(\omega_1)\propto e^{\kappa\omega_1}$, with $\kappa$ an arbitrary value that depends on the crystal orientation $\varphi$. We excluded the higher values of $\omega_1$ because the functional form of Eq.(\ref{sum_inv}) introduces significant errors within this frequency range. 

As a second comprobation, we analyze the validity of assigning $T_{NS}$ to the non-secular terms of Eq.(\ref{USMcompleto}). Then, we compute their effect on the time evolution of the density matrix of a simulated sample of 10 spins located at the sites of the $^1$H nuclei in a perfect lattice of CaSO$_4\cdot$2H$_2$O (it is worth to mention that the size of this small cluster suffices to calculate  the main frequency components the NMR spectra). As an example, the dipolar couplings for $\varphi=0º$ are listed  in Table II, and were calculated using the geometrical information from Ref. \cite{boeyens2002gypsum}. We simulate the experiment by calculating the density matrix $\sigma(t)$ at time $t = t_A+t_B$ (see Fig.\ref{MEsequence}), and the corresponding reverted signal amplitude  $\langle I_x \rangle = \Tr{ I_x \sigma(t)} $, in a closed system which first evolves under the secular dipolar Hamiltonian, and then under the ME propagator of Eq.(\ref{USMcompleto}) according to the Liouville equation, for the different angles $\varphi$. In this way, we evaluate the efficiency of the reversion sequence, when the only source of attenuation is the evolution under the non-reverted terms of the dipolar Hamiltonian (non-secular terms in Eq.(\ref{USMcompleto})), which are weighted by $\omega_1$. Though the calculated signal exhibits a complex dependence on the total reversion time, its overall shape is an exponential decay with characteristic  time $T_{NScalc}(\omega_1,\omega_D)$ which depends on the irradiation amplitude and the dipolar frequency.  

\begin{table}[]
	\centering

	\begin{tabular}{|c|c|c|c|c|c|c|c|c|c|c|}
		\cline{1-3} \cline{5-7}\cline{9-11}
		$k$ & $j$ & $\omega_D^{kj}$ ($\frac{\mathrm{KHz}}{6\pi}$) & &$k$ & $j$ & $\omega_D^{kj} $ ($\frac{\mathrm{KHz}}{6\pi}$)& &$k$ & $j$ & $\omega_D^{kj}$ ($\frac{\mathrm{KHz}}{6\pi}$) \\ \cline{1-3} \cline{5-7}\cline{9-11}
		1 & 2  & 12.71 & & 6 & 10 & -0.59 & &4 & 6  & 0.14 \\ \cline{1-3} \cline{5-7}\cline{9-11}
		3 & 4  & 12.71 & &7 & 10 & -0.49 & &3 & 8  & 0.14 \\ \cline{1-3} \cline{5-7}\cline{9-11}
		5 & 6  & 12.71 & &5 & 10 & -0.48 & &4 & 7  & 0.14 \\ \cline{1-3} \cline{5-7}\cline{9-11}
		7 & 8  & 12.71 & &6 & 9  & -0.48 & &1 & 8  & 0.12 \\ \cline{1-3} \cline{5-7}\cline{9-11}
		9 & 10 & 12.71 & & 7 & 9  & 0.42  & &4 & 5  & 0.12 \\ \cline{1-3} \cline{5-7}\cline{9-11}
		8 & 10 & -2.34 & &4 & 9  & 0.41  & &1 & 10 & 0.11 \\ \cline{1-3} \cline{5-7}\cline{9-11}
		2 & 4  & -2.08 & &3 & 9  & 0.27  & &2 & 7  & 0.09 \\ \cline{1-3} \cline{5-7}\cline{9-11}
		6 & 8  & -2.08 & &2 & 9  & 0.25  & &3 & 6  & 0.09 \\ \cline{1-3} \cline{5-7}\cline{9-11}
		1 & 3  & -1.98 & &4 & 10 & 0.23  & &3 & 7  & 0.09 \\ \cline{1-3} \cline{5-7}\cline{9-11}
		5 & 7  & -1.98 & &4 & 8  & 0.22  & &1 & 7  & 0.08 \\ \cline{1-3} \cline{5-7}\cline{9-11}
		1 & 4  & -1.21 & &5 & 9  & 0.18  & &3 & 5  & 0.08 \\ \cline{1-3} \cline{5-7}\cline{9-11}
		2 & 3  & -1.21 & &3 & 10 & 0.18  & &1 & 6  & 0.05 \\ \cline{1-3} \cline{5-7}\cline{9-11}
		5 & 8  & -1.21 & &1 & 9  & 0.17  & &2 & 5  & 0.05 \\ \cline{1-3} \cline{5-7}\cline{9-11}
		6 & 7  & -1.21 & &2 & 10 & 0.17  & &2 & 6  & 0.05 \\ \cline{1-3} \cline{5-7}\cline{9-11}
		8 & 9  & 1.01  & &2 & 8  & 0.14  & &1 & 5  & 0.05 \\ \cline{1-3} \cline{5-7}\cline{9-11}
	\end{tabular}
	\caption{Dipolar couplings of the simulated sample of 10 spins, for the case $\varphi=0$. }
\end{table}

\begin{figure}
\includegraphics[width=7.62cm,height=6.3cm]{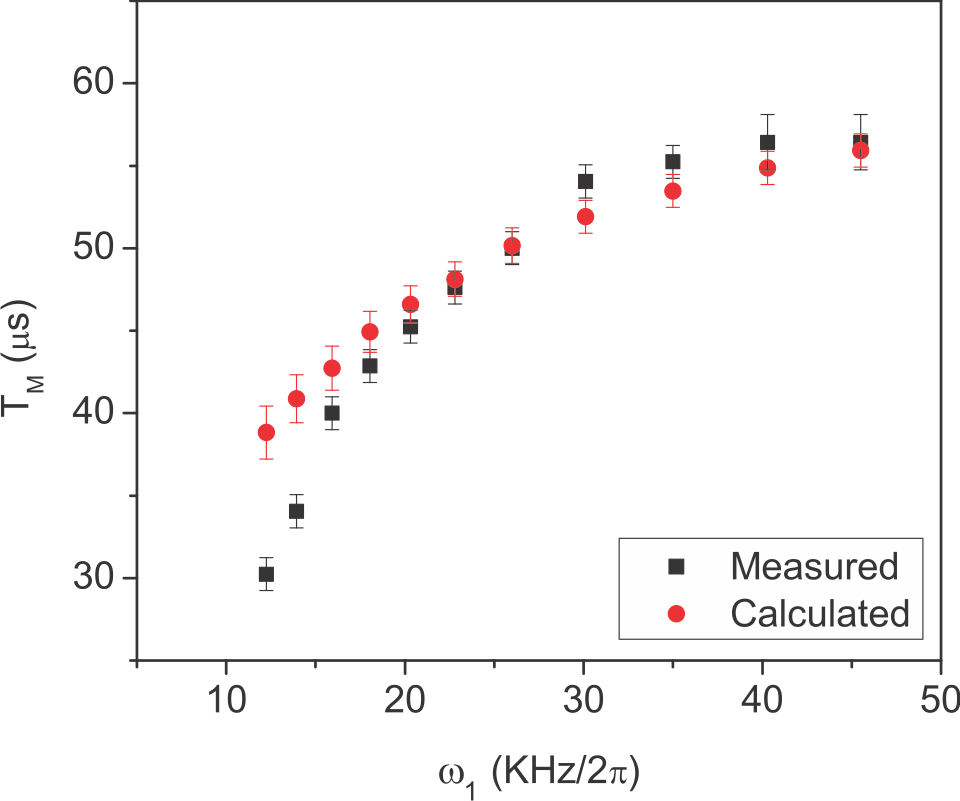}
\caption{Comparison between experimental and calculated values of $T_M = (T_{NScalc}^{-1}+\tau_D^{-1})^{-1}$, with $T_{NScalc}$ calculated on a ten-spin sample, by setting $\tau_D=61\mu$s.}
\label{comp_tm_ang0}
\end{figure}

 It is in principle possible to reproduce the measured data $T_M(\omega_1)$ by adding to the calculated rate $1/T_{NScalc}$ an extra contribution from decoherence. The obtained values for $\varphi=0º$, plotted in Fig.\ref{comp_tm_ang0}, are remarkably similar to the experiment, even though the contribution $T_{NScalc}$ was calculated in a small crystal of only ten spins. The slight discrepancy between $\tau_D=61\mathrm{\mu s}$ used in  Fig.\ref{comp_tm_ang0} and the experimental value $\tau_D=56\mathrm{\mu s}$ shown in Fig.\ref{tau-wD} suggests either that the measured value may not correspond to the actual plateau, which could only be achieved by increasing the pulse intensity $\omega_1$, or that $T_{NScalc}$ should be enhanced by computing the contribution of more spins. The small-size cluster may also explain the lower accuracy of the simulation, observed at $\varphi \neq 0$. However, the agreement found on the magnitudes and on the dependence with $\omega_1$, strongly supports hypothesis (\ref{sum_inv}). 

\subsection{Reverted signal amplitude in an Open quantum system} \label{signal}
In this section we explore the relation between the experimental times $\tau_D$ and the decoherence processes induced by the coupling of the observed system with an environment of harmonic phonons (irreversible adiabatic decoherence). In a recent work \cite{dominguez2016irreversible} we studied the adiabatic quantum decoherence in a system of interacting spins, coupled with a phonon environment, in the framework of the theory of open quantum systems. The observed system is a network of weakly interacting spin pairs; the bath corresponds to lattice phonons, and the system-environment interaction is generated by the variation of the dipole-dipole energy due to correlated shifts of the spin positions, produced by phonons. The model includes secular dipolar interaction between all spins in the sample, but the arrangement in pairs naturally ranks them, since the intra-pair interactions are larger than the inter-pair ones (provided that the pairs are sufficiently appart). Under the reasonable assumption that phonons are inefficient in producing transitions between spin levels, the time-evolution operator can be factorized as

\begin{equation}
U(t)=V_0(t) V(t) := e^{-it\H_S}e^{-it\left( \H_B+\H_I\right) } ,
\end{equation} 
where  $\H_S$ is the energy of the spin system, $\H_B$ is the phonon energy and $\H_I$ is the variation of the dipolar energy due to spin-phonon coupling. The coherence loss is determined by the decoherence function $\Gamma(t)$

\begin{equation}
\begin{split}
\sigma_{mn}(t) &= \sigma_{0mn}(t)e^{-\Gamma_{mn}(t)}= \\
&=\bra{m}V_0(t)\sigma(0)V_0^{\dagger}(t) \ket{n} \: \mathrm{Tr}_B\left\{ V_m(t)\rho_B\: V^{\dagger}_n(t)\right\}
\end{split} \label{dinamica1} 
\end{equation}
where $\ket{m}$ y $\ket{n}$ are elements of an uncoupled-pairs spin-basis $ \ket{m} = \ket{m_1,m_2,...,m_N}$, with $m_k$ the eigenvalue of the dipolar energy of the $k$th-pair. Additionally, it is assumed that only intra-pair fluctuations of dipolar energy are relevant, since inter-pair interactions are weaker. In this way, the decoherence function takes the form

\begin{equation}\label{gamma} 
\Gamma_{mn}(t)=\frac{ a K_B T }{\mu c^3} M^2\omega_D^2 t\approx 10^{-13} M^2 \omega_D^2 t
\end{equation}
where $\omega_D$ is the intrapair dipolar interaction, $a$ is the distance between adjacent pairs, $K_B$ the Boltzmann constant, $T$ the absolute temperature, $\mu$ the mass of the atom bearing the observed spin, $c$ the speed of sound in the sample and $M$ is the number of active pairs involved in the transition $\ket{m}=\ket{m_1,...,m_N}\rightarrow\ket{n}=\ket{n_1,...,n_N}$ (the pair $j$ is active if $m_j\neq n_j$). By setting the values $a=1$ nm, $T=300$ K, $\mu=1.66 10^{-27}$ kg, and $c=3000$ {m/s}, the decoherence function takes the form of the right term of Eq.(\ref{gamma}). This is a simplified expression of the decoherence function, that considers phonons propagating only along the direction of the intrapair interaction, that is, those which perturb the dipolar coupling more efficiently. The equivalent 3D function has a more complex expression that also is a decreasing function with a similar decay rate.

In order to explore the implication of the decoherence model of  Eq.(\ref{gamma}) on the observed amplitudes of the reverted NMR signal, it would be necessary to calculate the expectation value $\langle I_x\rangle$  using the reduced density matrix of  Eq.(\ref{dinamica1}). 
Even that  Eq.(\ref{gamma}) provides the decay rate of each element, the calculation of all the terms involved in the observable signal is still not available. Besides,  Eq.(\ref{gamma}) is valid during free evolution, but there is not an analogue under rf-irradiation, that is, during the reversion block of the ME sequence. 

However, the theoretical analysis can be carried out in terms of the system {\it purity}, $\mathcal{P}=\mathrm{Tr}(\sigma^2)$. This quantity is particularly useful because  its variation can be interpreted as a quantifier of environment induced decoherence \cite{ZurekHabibPaz93}, and can be estimated under reasonable hypotheses as shown below. At this point it is convenient to recall that the expectation value of any normalized operator $<\mathcal{O}>$ is bounded for the square root of the purity because the trace is an inner product in the space of the square complex matrices. Then, using the Cauchy-Schwarz inequality, yields

\begin{equation}
<\mathcal{O}>=\mathrm{Tr}(\sigma \mathcal{O}) \leq \abs{\sigma}\abs{\mathcal{O}} = \sqrt{\mathrm{Tr}(\sigma^2)} =\sqrt{\mathcal{P}}
\end{equation}
provided that $\abs{\mathcal{O}}=\sqrt{\mathrm{Tr}(\mathcal{O}^2)}=1$. We can also safely assume that the purity of the spins state do not increase while the system is being irradiated, because the rf pulses does not act on the lattice variables. So, we conclude that the purity just before the reversion block start is a good upper-bound for the square of the maximum reverted signal. We can now study the dependence of the phonon decoherence with the intrapair dipolar frequency $\omega_D$ and temperature $T$ through the behaviour of the purity
\begin{equation}  \label{pureza_1}
\mathcal{P}(t)=\sum_{mn}  \left| \sigma_{0mn}\right|^2 e^{-2\Gamma_{mn}(t)}.
\end{equation}
Each term of Eq.(\ref{pureza_1}) involves a characteristic number of active pairs $M$, and $\Gamma$ depends on ${m,n}$ only through the value of $M$. So, we can rewrite $\mathcal{P}$ as

\begin{equation}\label{pureza_sum}
\mathcal{P}(t)=\sum_{M} e^{-2\Gamma_{M}(t)} \left( \sum_{\left\lbrace mn\right\rbrace \in \mathcal{M}}  \left| \sigma_{0mn}\right|^2\right) ,
\end{equation}
where the index $\left\lbrace mn\right\rbrace \in \mathcal{M}$ means that the sum  involves states $\left\lbrace m,n\right\rbrace $ having $M$ active pairs.

The main obstacle in calculating  $g(M):=\sum_{\left\lbrace mn\right\rbrace \in \mathcal{M}}  \left| \sigma_{0mn}\right|^2$ is the half-knowledge of the closed, many-spin dynamics, encoded in $\sigma_0$. After the first pulse in Fig. \ref{MEsequence}, the state $\sigma_0=I_x$ is a statistical mixture of single-spin, single-quantum states. The subsequent evolution $V_0$  under the secular dipolar Hamiltonian does not increase the coherence order, while its flip-flop term is actually capable of increasing the number of active spins. Thus an estimation of the number of active spins is needed, but a a method for a direct calculation of this number is not available.

Inspired on the works by  Cho {\it et.al} \cite{cho2005multispin}, and Levy and Gleason \cite{levy1992multiple} (see Appendix), we represent $g(M)$ by a gaussian distribution 
\begin{equation}
g(M)=\frac{1}{\Delta M}e^{-\frac{\left( M-M_0(t)\right) ^2}{2\Delta M^2}}.
\end{equation} 	
whose center increases exponentially with time as

\begin{equation}\label{m0}
M_0(t)=e^{R\sqrt{M_2}t},
\end{equation}
with $R$ an arbitrary constant and $M_2$ the second moment of the crystal. The width of the distribution grows with $M_0$ with a factor of proportionality $q$, 

\begin{equation}\label{deltam}
\Delta M =qM_0(t)
\end{equation}

\begin{figure}
	\includegraphics[width=7.62cm,height=5.97cm]{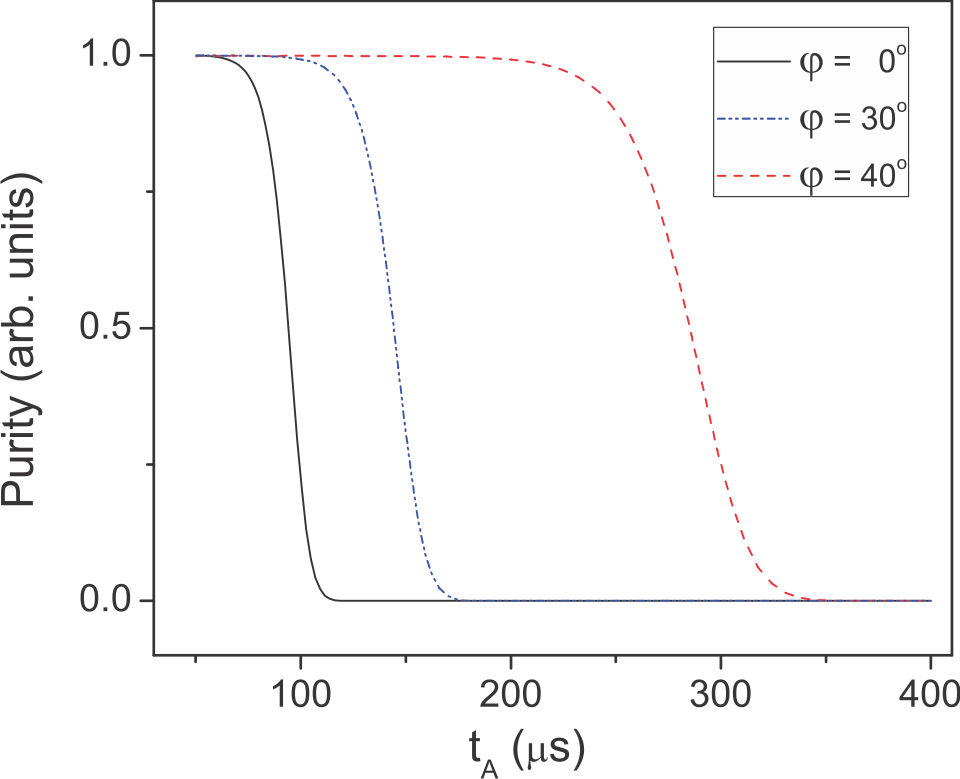}
	\caption{Purity of the spin pairs system as a function of the free evolution time, at three different crystal orientation angles.}
	\label{pureza}
\end{figure}

Then, in the continuum limit
\begin{equation}
\begin{array}{rl}
\mathcal{P}(t) &\propto \int_{0}^{\infty} e^{-\frac{2K_BTa\omega_D^2M^2t}{mc^3}}g(M)dM\\\\
		& \propto \int_{0}^{\infty} e^{-\frac{2K_BTa\omega_D^2M^2t}{mc^3}}e^{-\frac{\left( M-M_0(t)\right) ^2}{2\Delta M^2}}\frac{dM}{\Delta M}.
\end{array} \label{purezacontinuo}
\end{equation}
This expression gives a computable expression of the time dependence of the purity of a system of spin pairs coupled to a phonon bath. 
By performing a numerical integration, using a trapezoidal rule, we find that  $\mathcal{P}(t)$  is a decreasing sigmoid function that decays at different rates according to the crystal orientation as shown in Fig.\ref{pureza}. The different curves of purity correspond to  the dipolar frequencies  $\omega_D$ and second moments from  Table I. We set $R=2$ and $\Delta M=0.2M_0$ arbitrarily, to get insight in the dependence of $\mathcal{P}$ with $t$. 

\begin{figure}
	\includegraphics[width=9.05cm,height=12cm]{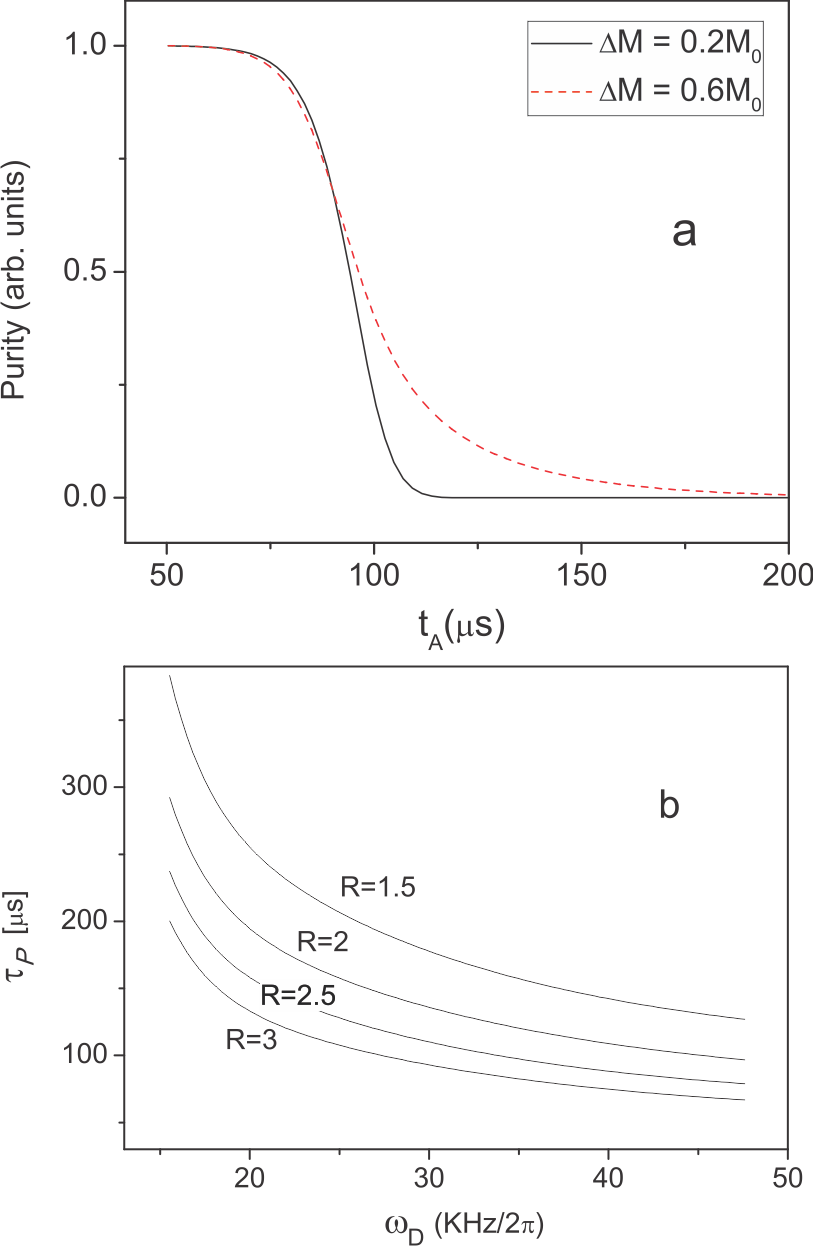}
	\caption{(a) Time dependence of purity calculated from Eq.(\ref{purezacontinuo}) with $q$= 0.2 (solid) and $q$= 0.6 (red dash). Both curves have very similar decay time. (b) Frequency dependence of the calculated decay time $\tau_P(\omega_{D})$ for different values of $R$. }
	\label{pureza_comportamiento}
\end{figure}

In order to give a qualitative description of the purity decay, let us characterize this decreasing function by the time $\tau_P$ at which the curve reaches the value $1/e$. The dependence of $\tau_P$ on the arbitrary parameter $q$  from Eq.(\ref{deltam}) is rather weak, as shown in Fig.(\ref{pureza_comportamiento}) (a) where the solid and dashed purity curves have similar characteristic times even when they correspond to very different values of $q$.  Figure (\ref{pureza_comportamiento}) (b) shows that $\tau_P$ has a different although moderate sensitivity to the parameter  $R$ from Eq.(\ref{m0}) at different dipolar frequencies (angles). 

\begin{figure}
	\includegraphics[width=7.65cm,height=6.1cm]{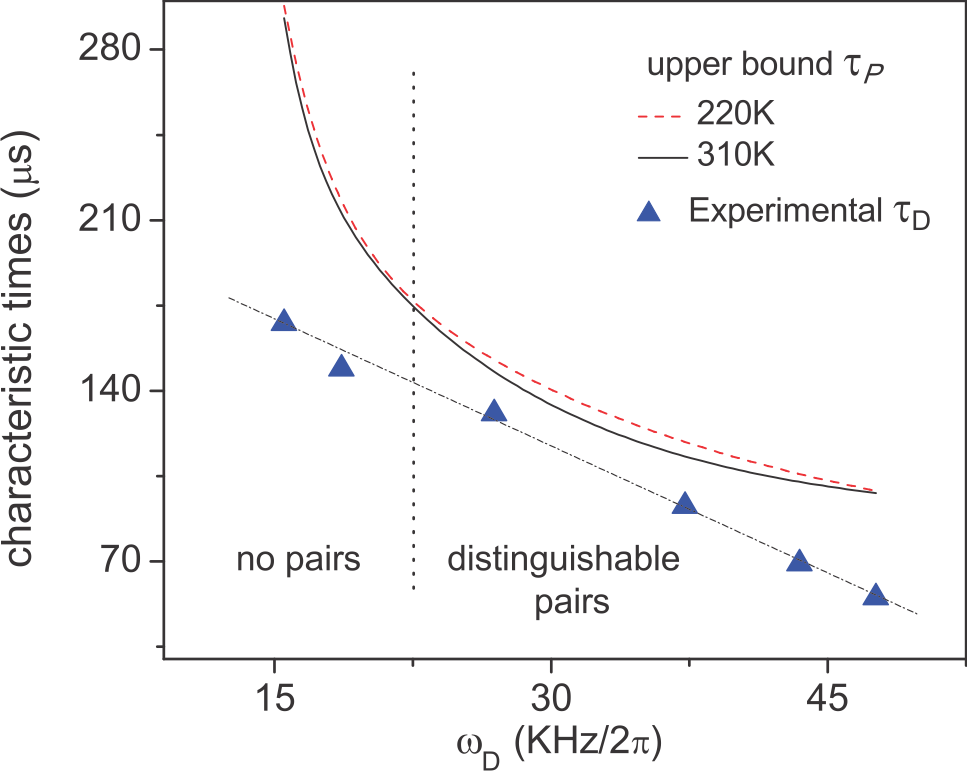}
	\caption{Experimental decoherence times (blue triangles) measured at different values of the intrapair dipolar coupling. The solid (black) and dashed (red) curves are the higher bound for the decoherence times calculated with the pair-boson model for decoherence. The model, as well as the experiment are insensible to a change of 100K in the sample temperature. The vertical line divides the regions where the NMR spectrum indicates the presence (or not) of well separated intrapair and interpair dipolar interactions.}
	\label{td_exp_calc}
\end{figure}

Besides this, it is worth to emphasize that the calculated values of  $\tau_P$ yielded by the theoretical prediction, have the same order of magnitude of the experimental ones, even when formula (\ref{pureza_sum}) corresponds to a simplified one-dimensional model. Figure \ref{td_exp_calc} shows that $\tau_P$ is in fact an upper bound for the measured $\tau_D$ (triangles), and that the purity decay time decreases with increasing dipolar frequency $\omega_{D}$ within the whole frequency range. The solid curves in the figure were calculated by setting $R=2$ and $q=0.2$. A result worth to remark is that the $\tau_P(\omega_{D})$ curves are almost insensible to a temperature change of 100 K, as shown by the lines at $T=220$ K (solid) and  $T=310$ K (dashed), which is in complete agreement with the experimental results reported in Sec. \ref{expresult}. 
It can be seen in Fig. \ref{td_exp_calc} that the  pair-boson model and the measured $\tau_D$ have a similar frequency dependence for $\omega_{D} > 20$kHz. In the lower frequency instead, the model underestimates the mechanisms that determine the experimental decoherence times. It should be noticed that this low frequency break down is consistent with the loss of the doublet shape in the NMR spectrum at the corresponding crystal orientations ($\varphi= 35^o$ and $40^o$). In other words, the pair-boson model may not describe the experiment because the spin system is actually not composed by easily distinguishable spin pairs.
In summary, we interpret the results of Fig. \ref{td_exp_calc} as an indication that the  intrinsic mechanism which attenuates the amplitude of the reverted signals may be attributed to the adiabatic decoherence due to the coupling of a system of interacting pairs with a phonon bath. This subtle correlation, although inefficient as a mechanism of relaxation, can alternatively be the mechanism which drives adiabatic decoherence.

\section{conclusions}

We studied the proton NMR signal amplitude response of a crystalline solid in reversion experiments. We find that the signal decay can be explained in terms of two main independent sources. One associated with the evolution under the non-secular terms of the spin interactions that cannot be reverted in the  experiment. This limitation can be mitigated to a high degree by performing the experiment under high field conditions in the rotating frame. The other source of signal reduction has a fundamental character, and was ascribed to environment induced decoherence.

 We measured the  characteristic attenuation times of the reverted  signals under different experimental conditions. The data are well described by two  contributions, one ($T_{NS}$) that depends on both the intensity of the rf pulses in the reversion sequence, $\omega_1$, and $\omega_{D}$, and one ($\tau_D$) which depends only on  $\omega_{D}$. The salient aspects are that $\tau_D$ is (a)  markedly sensible to changes in the intra-pair dipolar frequency, and (b) independent on the sample temperature within the probed range $220  - 310$K.  This behaviour led us to propose that the irreversible decay $\tau_D$ originates in the adiabatic loss of  quantum coherence mediated by the coupling of the dipole energy with low frequency phonons.

With the aim of probing the microscopic mechanism underlying the signal  decay, we calculate the reduced density matrix elements within the theory of open quantum systems. 
We estimated the spin system purity (which is an upper bound for the reverted signal amplitude) using the pair-phonon interaction model in its  1D version \cite{dominguez2016irreversible}. The model was supplemented with a statistical hypothesis to introduce the growth  of the spin cluster.
The resulting purity decay time $\tau_P(\omega_{D})$ showed a strong dependence with the dipolar frequency.
Its frequency dependence  is similar to that of the signal within a range of crystal orientation angles where the sample can safely be regarded as a linear chain of weakly interacting pairs, in contrast, it deviates from the experimental trend for orientation angles where the separation into stronger and weaker dipolar couplings no longer apply.
Within the region of discrepancy, the theoretical model underestimates the effectiveness of decoherence, revealing that the used model does not contemplate all the relevant mechanisms.
Anyway, this simple model of pairs allowed us to show up a frequency dependence which is a signature of the quantum adiabatic decoherence.
Likewise, the calculated purity is practically temperature independent. These features are in total consistency with the experimental characteristic times, which confirms that the observed irreversible decay can be explained by the adiabatic decoherence induced by the dipole-phonon coupling.

It is worth to remark  that our results involve quantitative estimations of the effect of decoherence in solid state NMR, starting from first principles and using a model Hamiltonian that represents many properties of a real sample. This kind of studies can faciliatate new  interesting applications related with vibrational properties of solids, growth of correlation in spin clusters, or protection of complex states against  environment induced degradation.

\section{Appendix}

Cho et.al. \cite{cho2005multispin}, studied the growth of coherence orders on the $x$-basis, in a crystalline spin system. The spin state is described by the density operator $\sigma$, and evolves under $V_0(t)=e^{-i\mathcal{H}_0t}$, as in Eq.(\ref{dinamica1}). Rotation about $y$-axis (that is, changing to the $x$-basis) leaves the spin state as  
	$$
	R_y\sigma(t)R_y^{-1} = R_yV_0(t)R_y^{-1} R_y\sigma(t=0)R_y^{-1} R_yV_0(t)R_y^{-1}.
	$$ 
The evolution of the rotated state involves exciting higher coherences, because the rotated evolution operator  $R_yV_0(t)R_y^{-1}$ is similar to a coherence pumping-sequence. This growth is explained in the work by Levy and Gleason \cite{levy1992multiple}, where they showed that, under pumping-sequences, the most populated coherence order raises exponentially with time, with a rate that depens on the second moment of the crystal (as Eq.(\ref{m0})).

This view led us to assume that the growth of the active spin pairs under the influence of secular dipolar evolution in the state $\sigma(t) = V_0(t)\sigma(t=0)V_0(t)$, is similar to the growth of coherence order in a rotated state $R_y\sigma(t)R_y^{-1}$. 

\bibliography{refgyp}

\begin{thebibliography}{19}%
\makeatletter
\providecommand \@ifxundefined [1]{%
 \@ifx{#1\undefined}
}%
\providecommand \@ifnum [1]{%
 \ifnum #1\expandafter \@firstoftwo
 \else \expandafter \@secondoftwo
 \fi
}%
\providecommand \@ifx [1]{%
 \ifx #1\expandafter \@firstoftwo
 \else \expandafter \@secondoftwo
 \fi
}%
\providecommand \natexlab [1]{#1}%
\providecommand \enquote  [1]{``#1''}%
\providecommand \bibnamefont  [1]{#1}%
\providecommand \bibfnamefont [1]{#1}%
\providecommand \citenamefont [1]{#1}%
\providecommand \href@noop [0]{\@secondoftwo}%
\providecommand \href [0]{\begingroup \@sanitize@url \@href}%
\providecommand \@href[1]{\@@startlink{#1}\@@href}%
\providecommand \@@href[1]{\endgroup#1\@@endlink}%
\providecommand \@sanitize@url [0]{\catcode `\\12\catcode `\$12\catcode
  `\&12\catcode `\#12\catcode `\^12\catcode `\_12\catcode `\%12\relax}%
\providecommand \@@startlink[1]{}%
\providecommand \@@endlink[0]{}%
\providecommand \url  [0]{\begingroup\@sanitize@url \@url }%
\providecommand \@url [1]{\endgroup\@href {#1}{\urlprefix }}%
\providecommand \urlprefix  [0]{URL }%
\providecommand \Eprint [0]{\href }%
\providecommand \doibase [0]{http://dx.doi.org/}%
\providecommand \selectlanguage [0]{\@gobble}%
\providecommand \bibinfo  [0]{\@secondoftwo}%
\providecommand \bibfield  [0]{\@secondoftwo}%
\providecommand \translation [1]{[#1]}%
\providecommand \BibitemOpen [0]{}%
\providecommand \bibitemStop [0]{}%
\providecommand \bibitemNoStop [0]{.\EOS\space}%
\providecommand \EOS [0]{\spacefactor3000\relax}%
\providecommand \BibitemShut  [1]{\csname bibitem#1\endcsname}%
\let\auto@bib@innerbib\@empty
\bibitem [{\citenamefont {Joos}(2003)}]{joos2003decoherence}%
  \BibitemOpen
  \bibfield  {author} {\bibinfo {author} {\bibfnamefont {E.}~\bibnamefont
  {Joos}},\ }\href@noop {} {\emph {\bibinfo {title} {Decoherence and the
  Appearance of a Classical World in Quantum Theory}}},\ Physics and astronomy
  online library\ (\bibinfo  {publisher} {Springer},\ \bibinfo {year}
  {2003})\BibitemShut {NoStop}%
\bibitem [{\citenamefont {Zurek}(2003)}]{zurek03}%
  \BibitemOpen
  \bibfield  {author} {\bibinfo {author} {\bibfnamefont {W.}~\bibnamefont
  {Zurek}},\ }\href@noop {} {\bibfield  {journal} {\bibinfo  {journal} {Rev.
  Mod. Phys.}\ }\textbf {\bibinfo {volume} {75}},\ \bibinfo {pages} {715}
  (\bibinfo {year} {2003})}\BibitemShut {NoStop}%
\bibitem [{\citenamefont {Yu}\ and\ \citenamefont
  {Eberly}(2002)}]{yu-eberly02}%
  \BibitemOpen
  \bibfield  {author} {\bibinfo {author} {\bibfnamefont {T.}~\bibnamefont
  {Yu}}\ and\ \bibinfo {author} {\bibfnamefont {J.~H.}\ \bibnamefont
  {Eberly}},\ }\href {\doibase 10.1103/PhysRevB.66.193306} {\bibfield
  {journal} {\bibinfo  {journal} {Phys. Rev. B}\ }\textbf {\bibinfo {volume}
  {66}},\ \bibinfo {pages} {193306} (\bibinfo {year} {2002})}\BibitemShut
  {NoStop}%
\bibitem [{\citenamefont {Petitjean}\ and\ \citenamefont
  {Jacquod}(2006)}]{PetitjeanJacq06}%
  \BibitemOpen
  \bibfield  {author} {\bibinfo {author} {\bibfnamefont {C.}~\bibnamefont
  {Petitjean}}\ and\ \bibinfo {author} {\bibfnamefont {P.}~\bibnamefont
  {Jacquod}},\ }\href {\doibase 10.1103/PhysRevLett.97.124103} {\bibfield
  {journal} {\bibinfo  {journal} {Phys. Rev. Lett.}\ }\textbf {\bibinfo
  {volume} {97}},\ \bibinfo {pages} {124103} (\bibinfo {year}
  {2006})}\BibitemShut {NoStop}%
\bibitem [{\citenamefont {González}\ \emph {et~al.}(2011)\citenamefont
  {González}, \citenamefont {Segnorile},\ and\ \citenamefont
  {Zamar}}]{GzzSegZam11}%
  \BibitemOpen
  \bibfield  {author} {\bibinfo {author} {\bibfnamefont {C.~E.}\ \bibnamefont
  {González}}, \bibinfo {author} {\bibfnamefont {H.~H.}\ \bibnamefont
  {Segnorile}}, \ and\ \bibinfo {author} {\bibfnamefont {R.~C.}\ \bibnamefont
  {Zamar}},\ }\href@noop {} {\bibfield  {journal} {\bibinfo  {journal} {Phys.
  Rev. E}\ }\textbf {\bibinfo {volume} {83}},\ \bibinfo {pages} {011705}
  (\bibinfo {year} {2011})}\BibitemShut {NoStop}%
\bibitem [{\citenamefont {Segnorile}\ and\ \citenamefont
  {Zamar}(2011)}]{SegZam11}%
  \BibitemOpen
  \bibfield  {author} {\bibinfo {author} {\bibfnamefont {H.}~\bibnamefont
  {Segnorile}}\ and\ \bibinfo {author} {\bibfnamefont {R.}~\bibnamefont
  {Zamar}},\ }\href {\doibase http://dx.doi.org/10.1063/1.3668559} {\bibfield
  {journal} {\bibinfo  {journal} {The Journal of Chemical Physics}\ }\textbf
  {\bibinfo {volume} {135}},\ \bibinfo {pages} {244509} (\bibinfo {year}
  {2011})}\BibitemShut {NoStop}%
\bibitem [{\citenamefont {Morgan}\ \emph {et~al.}(2012)\citenamefont {Morgan},
  \citenamefont {Oganesyan},\ and\ \citenamefont
  {Boutis}}]{morgan2012multispin}%
  \BibitemOpen
  \bibfield  {author} {\bibinfo {author} {\bibfnamefont {S.~W.}\ \bibnamefont
  {Morgan}}, \bibinfo {author} {\bibfnamefont {V.}~\bibnamefont {Oganesyan}}, \
  and\ \bibinfo {author} {\bibfnamefont {G.~S.}\ \bibnamefont {Boutis}},\
  }\href@noop {} {\bibfield  {journal} {\bibinfo  {journal} {Physical Review
  B}\ }\textbf {\bibinfo {volume} {86}},\ \bibinfo {pages} {214410} (\bibinfo
  {year} {2012})}\BibitemShut {NoStop}%
\bibitem [{\citenamefont {Mozyrski}\ and\ \citenamefont
  {Privman}(1988)}]{priv98}%
  \BibitemOpen
  \bibfield  {author} {\bibinfo {author} {\bibfnamefont {D.}~\bibnamefont
  {Mozyrski}}\ and\ \bibinfo {author} {\bibfnamefont {V.}~\bibnamefont
  {Privman}},\ }\href@noop {} {\bibfield  {journal} {\bibinfo  {journal} {J.
  Stat. Phys.}\ }\textbf {\bibinfo {volume} {91}},\ \bibinfo {pages} {787}
  (\bibinfo {year} {1988})}\BibitemShut {NoStop}%
\bibitem [{\citenamefont {Costa}\ \emph {et~al.}(2016)\citenamefont {Costa},
  \citenamefont {Beims},\ and\ \citenamefont {Strunz}}]{strunz16}%
  \BibitemOpen
  \bibfield  {author} {\bibinfo {author} {\bibfnamefont {A.~C.~S.}\
  \bibnamefont {Costa}}, \bibinfo {author} {\bibfnamefont {M.~W.}\ \bibnamefont
  {Beims}}, \ and\ \bibinfo {author} {\bibfnamefont {W.~T.}\ \bibnamefont
  {Strunz}},\ }\href {\doibase 10.1103/PhysRevA.93.052316} {\bibfield
  {journal} {\bibinfo  {journal} {Phys. Rev. A}\ }\textbf {\bibinfo {volume}
  {93}},\ \bibinfo {pages} {052316} (\bibinfo {year} {2016})}\BibitemShut
  {NoStop}%
\bibitem [{\citenamefont {Borrelli}\ \emph {et~al.}(2013)\citenamefont
  {Borrelli}, \citenamefont {Haikka}, \citenamefont {De~Chiara},\ and\
  \citenamefont {Maniscalco}}]{haikka2013}%
  \BibitemOpen
  \bibfield  {author} {\bibinfo {author} {\bibfnamefont {M.}~\bibnamefont
  {Borrelli}}, \bibinfo {author} {\bibfnamefont {P.}~\bibnamefont {Haikka}},
  \bibinfo {author} {\bibfnamefont {G.}~\bibnamefont {De~Chiara}}, \ and\
  \bibinfo {author} {\bibfnamefont {S.}~\bibnamefont {Maniscalco}},\ }\href
  {\doibase 10.1103/PhysRevA.88.010101} {\bibfield  {journal} {\bibinfo
  {journal} {Phys. Rev. A}\ }\textbf {\bibinfo {volume} {88}},\ \bibinfo
  {pages} {010101} (\bibinfo {year} {2013})}\BibitemShut {NoStop}%
\bibitem [{\citenamefont {Dom{\'\i}nguez}\ \emph {et~al.}(2016)\citenamefont
  {Dom{\'\i}nguez}, \citenamefont {Gonz{\'a}lez}, \citenamefont {Segnorile},\
  and\ \citenamefont {Zamar}}]{dominguez2016irreversible}%
  \BibitemOpen
  \bibfield  {author} {\bibinfo {author} {\bibfnamefont {F.}~\bibnamefont
  {Dom{\'\i}nguez}}, \bibinfo {author} {\bibfnamefont {C.}~\bibnamefont
  {Gonz{\'a}lez}}, \bibinfo {author} {\bibfnamefont {H.}~\bibnamefont
  {Segnorile}}, \ and\ \bibinfo {author} {\bibfnamefont {R.}~\bibnamefont
  {Zamar}},\ }\href@noop {} {\bibfield  {journal} {\bibinfo  {journal}
  {Physical Review A}\ }\textbf {\bibinfo {volume} {93}},\ \bibinfo {pages}
  {022120} (\bibinfo {year} {2016})}\BibitemShut {NoStop}%
\bibitem [{\citenamefont {Zurek}\ \emph {et~al.}(1993)\citenamefont {Zurek},
  \citenamefont {Habib},\ and\ \citenamefont {Paz}}]{ZurekHabibPaz93}%
  \BibitemOpen
  \bibfield  {author} {\bibinfo {author} {\bibfnamefont {W.~H.}\ \bibnamefont
  {Zurek}}, \bibinfo {author} {\bibfnamefont {S.}~\bibnamefont {Habib}}, \ and\
  \bibinfo {author} {\bibfnamefont {J.~P.}\ \bibnamefont {Paz}},\ }\href
  {\doibase 10.1103/PhysRevLett.70.1187} {\bibfield  {journal} {\bibinfo
  {journal} {Phys. Rev. Lett.}\ }\textbf {\bibinfo {volume} {70}},\ \bibinfo
  {pages} {1187} (\bibinfo {year} {1993})}\BibitemShut {NoStop}%
\bibitem [{\citenamefont {Boeyens}\ and\ \citenamefont
  {Ichharam}(2002)}]{boeyens2002gypsum}%
  \BibitemOpen
  \bibfield  {author} {\bibinfo {author} {\bibfnamefont {J.~C.}\ \bibnamefont
  {Boeyens}}\ and\ \bibinfo {author} {\bibfnamefont {V.}~\bibnamefont
  {Ichharam}},\ }\href@noop {} {\bibfield  {journal} {\bibinfo  {journal}
  {Zeitschrift f{\"u}r Kristallographie-New Crystal Structures}\ }\textbf
  {\bibinfo {volume} {217}},\ \bibinfo {pages} {9} (\bibinfo {year}
  {2002})}\BibitemShut {NoStop}%
\bibitem [{\citenamefont {Pake}(1948)}]{pake48}%
  \BibitemOpen
  \bibfield  {author} {\bibinfo {author} {\bibfnamefont {G.~E.}\ \bibnamefont
  {Pake}},\ }\href@noop {} {\bibfield  {journal} {\bibinfo  {journal} {The
  Journal of chemical physics}\ }\textbf {\bibinfo {volume} {16}},\ \bibinfo
  {pages} {327} (\bibinfo {year} {1948})}\BibitemShut {NoStop}%
\bibitem [{\citenamefont {Keller}(1988)}]{keller88}%
  \BibitemOpen
  \bibfield  {author} {\bibinfo {author} {\bibfnamefont {A.}~\bibnamefont
  {Keller}}\ }(\bibinfo  {publisher} {Academic Press},\ \bibinfo {year}
  {1988})\ pp.\ \bibinfo {pages} {183 -- 246}\BibitemShut {NoStop}%
\bibitem [{\citenamefont {Rhim}\ \emph {et~al.}(1971)\citenamefont {Rhim},
  \citenamefont {Pines},\ and\ \citenamefont {Waugh}}]{rpw71}%
  \BibitemOpen
  \bibfield  {author} {\bibinfo {author} {\bibfnamefont {W.-K.}\ \bibnamefont
  {Rhim}}, \bibinfo {author} {\bibfnamefont {A.}~\bibnamefont {Pines}}, \ and\
  \bibinfo {author} {\bibfnamefont {J.~S.}\ \bibnamefont {Waugh}},\ }\href
  {\doibase 10.1103/PhysRevB.3.684} {\bibfield  {journal} {\bibinfo  {journal}
  {Phys. Rev. B}\ }\textbf {\bibinfo {volume} {3}},\ \bibinfo {pages} {684}
  (\bibinfo {year} {1971})}\BibitemShut {NoStop}%
\bibitem [{\citenamefont {Abragam}\ and\ \citenamefont
  {Goldman}(1982)}]{abragol82}%
  \BibitemOpen
  \bibfield  {author} {\bibinfo {author} {\bibfnamefont {A.}~\bibnamefont
  {Abragam}}\ and\ \bibinfo {author} {\bibfnamefont {M.}~\bibnamefont
  {Goldman}},\ }\href {http://books.google.com.ar/books?id=JMrvAAAAMAAJ} {\emph
  {\bibinfo {title} {Nuclear magnetism: order and disorder}}},\ International
  series of monographs on physics\ (\bibinfo  {publisher} {Clarendon Press},\
  \bibinfo {year} {1982})\BibitemShut {NoStop}%
\bibitem [{\citenamefont {Cho}\ \emph {et~al.}(2005)\citenamefont {Cho},
  \citenamefont {Ladd}, \citenamefont {Baugh}, \citenamefont {Cory},\ and\
  \citenamefont {Ramanathan}}]{cho2005multispin}%
  \BibitemOpen
  \bibfield  {author} {\bibinfo {author} {\bibfnamefont {H.}~\bibnamefont
  {Cho}}, \bibinfo {author} {\bibfnamefont {T.~D.}\ \bibnamefont {Ladd}},
  \bibinfo {author} {\bibfnamefont {J.}~\bibnamefont {Baugh}}, \bibinfo
  {author} {\bibfnamefont {D.~G.}\ \bibnamefont {Cory}}, \ and\ \bibinfo
  {author} {\bibfnamefont {C.}~\bibnamefont {Ramanathan}},\ }\href@noop {}
  {\bibfield  {journal} {\bibinfo  {journal} {Physical Review B}\ }\textbf
  {\bibinfo {volume} {72}},\ \bibinfo {pages} {054427} (\bibinfo {year}
  {2005})}\BibitemShut {NoStop}%
\bibitem [{\citenamefont {Levy}\ and\ \citenamefont
  {Gleason}(1992)}]{levy1992multiple}%
  \BibitemOpen
  \bibfield  {author} {\bibinfo {author} {\bibfnamefont {D.}~\bibnamefont
  {Levy}}\ and\ \bibinfo {author} {\bibfnamefont {K.}~\bibnamefont {Gleason}},\
  }\href@noop {} {\bibfield  {journal} {\bibinfo  {journal} {The Journal of
  Physical Chemistry}\ }\textbf {\bibinfo {volume} {96}},\ \bibinfo {pages}
  {8125} (\bibinfo {year} {1992})}\BibitemShut {NoStop}%
\end{thebibliography}%
\bibliographystyle{apsrev4-1} 

\end{document}